\documentclass[12pt,preprint]{aastex}
\begin{document}
\title{Rotational Coupling of the Pinned Core Superfluid}
\author{M. Jahan-Miri}
\affil{Department of Physics, Shiraz University, Shiraz 71454,
Iran}
\authoremail{jahan@physics.susc.ac.ir}

\begin{abstract}
The effects of pinning between fluxoids and vortices in the core
of a neutron star, on the dynamics of the core neutron superfluid
are considered. The pinning impedes, but does not absolutely
block, any radial as well as {\em azimuthal} motion of the neutron
vortices with respect to the lattice of fluxoids. The time scale
for the coupling of rotation of the core superfluid to the rest of
the star is calculated, allowing for the effect of the finite
frictional force on the neutron vortices due to their pinning with
the fluxoids. This turns out to be the dominant mechanism for the
coupling of the core of a neutron star to its crust, as compared
to the role of electron scattering, for most cases of interest.
Furthermore, different behaviors for the post-glitch response of
the core superfluid are distinguished that might be tested against
the relevant observational data. Also, a conceptually important
case (and controversial too, in the earlier studies on the role of
the crustal superfluid) is realized where a superfluid may remain
decoupled in spite of an spinning up of its vortices.
\end{abstract}

\section{INTRODUCTION}
The neutron superfluid in the core of a neutron star coexists with
the super-conducting protons and the ``normal'' degenerate
electrons. The interior charged plasma including the lattice of
the proton vortices (fluxoids) is expected to be strongly coupled
to the lattice of nuclei and the electrons in the crust due to the
strong magnetic field present, with a coupling time scale $\leq
10$~s \citep{als84}.  All charged components of the star are hence
considered as one co-rotating ``component'', referred to as the
``crust''; in contrast to the neutron superfluid in the core,
being the second ``component''. An isolated neutron star is
subject to an electromagnetic spin-down torque that acts on its
magnetic field, hence on the ``crust''. The core superfluid would
be likewise driven to follow the long-term spinning down of the
crust through the mutual coupling mechanism that would operate
between the crust and the superfluid vortices. The force, on the
vortices, responsible for such a spinning down (or any assumed
short-term relaxation) of the core superfluid has been, generally,
considered to be only that of the scattering of the electrons off
the vortex cores. The dominant scattering effect is that of the
induced magnetization of the vortices caused by a ``drag'' between
the proton and neutron condensates.  The associated vortex
velocity relaxation time scale is $\tau_{\rm e} \sim 10 P_{\rm s}$
-- $20 P_{\rm s} $, where $P_{\rm s}$ is the spin period of the
star \citep{alet84}.

However, in the quantum liquid interior of a neutron star a
neutron vortex is expected to ``pin'' to a fluxoid should the two
structures overlap. The mechanism of the pinning is associated
with either the proton density perturbations or the magnetized
nature of both the  fluxoids as well as the neutron vortices. The
strength $E_{\rm P}$ of the corresponding energy barrier has been
estimated as $E_{\rm P} \sim 0.1$ or $1.0$~MeV, for the two
mechanisms, respectively. Likewise, the effective length scale
$d_{\rm P}$ of the pinning interaction is expected to be of the
order of the coherence length $\xi_{\rm p}$ or the London
penetration depth $\lambda_{\rm p}$ of the proton superconductor
for the above two mechanisms, respectively. The effective pinning
force (per intersection) $f_{\rm P} = {E_{\rm P} \over d_{\rm P}}
$ is therefore roughly the same for the two interaction
mechanisms, since the value of $E_{\rm P}$ due to the magnetic
interaction is larger than that of the density perturbation by the
same ratio as the inverse of their corresponding $d_{\rm P}$
values, ie.  ${ \lambda_{\rm p} \over \xi_{\rm p}} \sim 10$
\citep{mt85,sau89,jon91a,jon06}.

The consequences of such a pinning with respect to the {\em
radial} motion of the vortices have been discussed in the
literature, to some extent.  The outward moving vortices in a
slowing down neutron star have been argued to sweep the fluxoids
along with them and hence expel the magnetic flux out of the
stellar core \citep{sbmt90,jon91b,sedd92}. The magnetic evolution
of pulsars, particularly those in binary systems, has been
calculated, based on this effect
\citep{dcc93,jb94,mj96,ugk97,mj00}. Also, due to the steady-state
slowing down of the star, the superfluid would be rotating faster
than the vortices, resulting in a radial Magnus ``force'' that
drives them against the pinning barriers. Glitch inducing
mechanisms, caused by the core superfluid, have been suggested on
this ground, invoking the possibility of sudden disturbances
\citep{mt85}, and based on the predicted long-term evolution of
the steady-state value of the associated rotational lag between
the superfluid and the vortices \citep{mj02}. Also, the expected
post-glitch behavior of the observable rotation frequency of the
crust of a neutron star has been discussed in this context
\citep{ccd92}.

The pinning would however also impede any assumed relative
azimuthal motion between the two lattices of the vortex lines, and
impart {\em azimuthal} forces. That is, it could impart a torque,
and act as a means for a transfer of angular momentum to the core
superfluid. The effect is, in principle, same as the role of the
standard ``static'' frictional force, realized in the laboratory
experiments on a rotating superfluid, in presence of the pinning
centers \citep{tt80,adam85}. Such a role for the azimuthal
component of the pinning force in the core of a neutron star was
referred to by \citet{mj98}, and has been recently studied by
\citet{sa09}. These authors assume however absolute pinning, and
assign an {\em infinite} strength to the azimuthal component of
the pinning force, allowing only for a sliding motion of the
vortices along the floxoids, as was first suggested by
\citet{sau89}. It should be clarified, at the outset, that we are
concerned with only an application of the standard results and
basic formulation of the superfluid rotational dynamics, at the
hydrodynamical level \citep{son87}, avoiding complications due to
the microscopic details. We adopt the generally accepted standard
structure for the interior of a neutron star. We attempt to
include the role of the azimuthal force on the vortices, due to
their pinning with the fluxoids, in the rotational dynamics of the
core superfluid, in addition to the viscous drag force, due to the
scattering of the electrons.

In the following we discuss the rotational dynamics of neutron
stars taking into account also the above effects (both radial, as
well as azimuthal) due to the pinning of the vortices in the core
of the star.  The steady-state relative rotation of the different
components of a neutron star is briefly described, distinguishing
between the cases of {\em pinned} and {\em unpinned} vortices in
the core superfluid. It may be noted, that a clear picture of the
steady-state relative rotations is indeed needed in order to
follow the discussion further. The vortex-velocity relaxation time
scale is then calculated, allowing also for the finite effect of
the frictional force of the pinning with the fluxoids. The
dynamical timescale for the coupling of the pinned core superfluid
is thence derived. In addition, qualitatively different behaviors
for the response of the core superfluid to a jump in the rotation
frequency of the ``crust'' are distinguished, depending on the
magnitude of the jump, in the case of pinned superfluid. In the
final section, the relevance of our results to the observational
data on pulsars glitches are discussed briefly, indicating the
possibility of distinguishing among the different cases
considered.

\section{The CORE SUPERFLUID}
Superfluid vortices move with the local superfluid velocity except
when there is an external force acting on a vortex. For a given
external force $F_{\rm ex}$, per unit length of the vortex, the
equation of motion is given \citep{son87} as
\begin{eqnarray}
\vec{F}_{\rm ex} & = &   \rho_{\rm s} \vec{\kappa} \times
                (\vec{v}_{\rm s} - \vec{v}_{\rm L})
\end{eqnarray}
where $\vec{v_{\rm s}}$ and $\vec{v_{\rm L}}$ are  velocities of
the local superfluid and the vortex line, respectively, $\rho_{\rm
s}$ is the superfluid mass density, $\vec{\kappa}$ is the
vorticity of the vortex line directed along the rotation axis,
with a magnitude $\kappa = { h \over 2 m_{\rm n}}$ for the neutron
superfluid, where $m_{\rm n}$ is the mass of a neutron. The
kinematic side of the equation is generally referred to as the
Magnus "force''. The force, arising from the gradient of the
superfluid kinetic energy, is exerted between the superfluid and
the vortices, and explains how a ``mutual'' friction between the
superfluid and its environment is realized. As is usual in this
field, we will neglect the collective effects of the vortex
lattice, finite tension of a vortex, as well as any relevant
non-dissipative force on a vortex by the lattice, which has been
studied recently for the vortices in the crust of neutron stars
\citep{lin09}.

In the case of a  rotating superfluid, the number density $n_{\rm
v}$, per unit area, of the vortices obeys \(\kappa n_{\rm v} = 2
\Omega_{\rm s}\), and the rate $\dot \Omega_{\rm s}$ of change of
the rotation frequency $\Omega_{\rm s}$ of the superfluid is
associated with a radial velocity $v_r$ of the vortices, \( \dot
\Omega_{\rm s} = - 2 \ { \Omega_{\rm s} \over r} \ {v_r} \), where
$r$ is the distance from the rotation axis, and $v_r>0$ is in the
outward direction. Thus, a spinning down (up) torque on a rotating
superfluid is accompanied by a radial outward (inward) motion of
the vortices.

\subsection{Steady-State Relative Rotations}
During the steady state {\em spinning down} of a neutron star, the
vortices are expected to be co-rotating with the crust, including
the proton condensate and the lattice of fluxoids, in the core of
the star. Strict co-rotation of the vortices with the crust is not
however possible for an assumed vortex relaxation process due only
to the electron scattering. In contrast, the pinning force could
be imparted even while the vortices are co-rotating with the
fluxoids. The pinning potential barrier could act as a ``stretched
spring'', that does pull the mass attached to it, even while both
are moving together. In any case, the difference in the rotation
rate between the vortices and the fluxoids (the crust) is tiny and
will be neglected, for the steady state considered. On the other
hand, the steady-state rotation rate of the superfluid should be
different than that of the crust and/or the vortices, due to the
assumed pinning of the vortices. The superfluid rotational lag is
defined as $\omega = \Omega_{\rm s} -\Omega_{\rm L}$, where
$\Omega_{\rm L}$ is the rotation frequency of the neutron vortex
lines.

\subsubsection{The Steady-state Lag}
The superfluid critical lag $\omega_{\rm cr}$ is determined from a
balance of the radial Magnus force  \( {F_{\rm M}}_{\rm r} =
\rho_{\rm s} \kappa r \omega  \) with the radial component of the
external force, $F_{\rm r}$, on the vortices, per unit length.
Equating $F_{\rm r}$ with the pinning force results in
\citep{dcc93}
\begin{eqnarray}
 \omega_{\rm cr} & = & { f_{\rm P} \over \rho_{\rm s} \, \kappa \, r \, d_{\rm
 f}}\\
                 & \sim & 1.6 \times 10^{-4} \, ({\rm rad \ s}^{-1}) \ {B_{12}^{1/2} \over r_6},
\end{eqnarray}
where $r_6$ is the value of $r$ in units of $10^6$~cm, and
$B_{12}$ is the strength of the magnetic field $B_{\rm c}$ in the
stellar core in units of $10^{12}$~G. For the typical magnetic
fields of young pulsars, one finds typical values of
\begin{itemize}
\item[{\em i)}]  $ \omega_{\rm cr}  \gtrsim 10^{-4} {\ \rm rad \
s}^{-1}$,\\ for $R_6 \sim 1$, at the outer regions of the stellar
core, and \item[{\em ii)}] $ \omega_{\rm cr} \gtrsim 10^{-3} { \
\rm rad \ s}^{-1}$, \\ for $R_6 \sim 0.1$, at the inner parts,
embracing $ \sim 1 \% $ of the moment of inertia of the core.
\end{itemize}

The steady-state spinning down will be thus established while \(
\Omega_{\rm c} = \Omega_{\rm L} < \Omega_{\rm s} \), \(
\dot\Omega_{\rm s} = \dot\Omega_{\rm c} \), and the steady-state
vale of the lag \( \omega_\infty =  \omega_{\rm cr} > 0 \), with
the above estimates for its value (which also indicates a state of
differential rotation for the core superfluid at the different
radial distances). It may be noted that random unpinning of the
vortices could in principle serve to establish even smaller values
of the lag $\omega_\infty < \omega_{\rm cr}$. An estimate of the
effect, based on the vortex creep model, shows however the
difference to be negligible \citep{ccd92}. Thus, we will be
neglecting the role of random unpinning of the vortices, and hence
assuming $\omega_\infty = \omega_{\rm cr}$, in the following.

The expected relative rotation of the ``crust'', the core
superfluid, and its vortices is sketched in Fig.~1, as a visual
aid which turns out to be vital too. The sketch helps to keep
track and be ensured that the, preliminary but, fundamental
physics applicable to a transfer of angular momentum between the
superfluid and its environment (the crust, here) is obeyed! That
is, for a superfluid spin-down (-up) to be achieved the crust must
be, or tend to be, rotating slower (faster) than not only the
superfluid itself, but also slower (faster) than the vortices (see
\S 3.2.2 and \S 4, below). Each plot indicates the three
successive stages at a disturbance and the transitions between
them. Namely, the conditions during a steady state, as discussed
above (marked as ``S.S.''), followed by a jump in the rotation
frequency of the crust (``JUMP''), and after the core has
responded to that jump, returning to a new steady state
(``post-G''). The cases of pinned and free vortices behave
differently and are shown separately, as there is also a further
distinction for the pinned cases depending on the relative size of
the jump, to be discussed below.

\subsection{Vortex Velocity Relaxation}
An assumed departure from the steady-state co-rotation of the
vortices with the fluxoids brings about their continually crossing
one another, with an energy cost of the pinning energy $E_{\rm
P}$, per intersection, associated to a frictional force $f_{\rm
P}$. The total number $N_{\rm X}$ of crossings per relative
rotation cycle of the two lattices (embedded in a spherical
boundary surface) is $N_{\rm X} = N_{\rm f} N_{\rm v} \sin{\chi}$
(see the Appendix), where $N_{\rm f}$ and $N_{\rm v}$ are the
total number of the fluxoids and the vortices, and $\chi$ is the
angle between the two families of lines, namely the angle of
inclination between the rotation and the magnetic axes of the
star.  Excluding the cases of parallel lattices corresponding to
$\chi \sim 0 \ {\rm or} \ 180$~deg which are not common among the
observed pulsars, the correction due to the $\sin{\chi}$ factor
may be neglected, assuming near perpendicular geometry between the
two lattices, for simplicity.

The {\em effective} frictional force $F_{\rm pin}$ of the pinning,
per unit length of a vortex, while cutting through the lattice of
uniformly distributed fluxoids (with a spacing $d_{\rm f}$) may be
estimated by taking a {\em time average} over a cycle (``pinning
cycle'') of passing through a pinning region, of a length $l_1 =
d_{P}$, and an adjacent inter-pinning zone, of a length $l_2 =
d_{\rm f} - d_{P}$. This is in contrast to the drag force due to
the electron scattering $F_{\rm elec} $, which acts continuously,
and will be treated separately. Thus,
\begin{eqnarray}
F_{\rm pin}   =  {1 \over T_{\rm cyc} } (t_1 \, f_1 + t_2 \, f_2)
\end{eqnarray}
where $T_{\rm cyc} = t_1 + t_2 \ $ is the total time for a vortex
to pass through one pinning cycle, and the subscripts refer to the
two regions, respectively. The pinning force is operative only
within each pinning zone, namely
\begin{eqnarray}
f_1  & = &  N_{\rm c} \ f_{\rm P} \\
     & = &  {f_{\rm P} \over d_{\rm f}} = {E_{\rm P} \over
                                       d_{\rm P} \, d_{\rm f}}, \\
f_2  & = & 0,
\end{eqnarray}
where we have used $N_{\rm c} = \frac{1}{d_{\rm f}}$ for the
number $N_{\rm c}$ of the potential crossings per unit length of a
vortex, per pinning cycle (see the Appendix).

Next, the microscopic motion of any given vortex, while cutting
through the fluxoids, might be expected to fall in between the
following two limiting types of behaviors, for a first
approximation.
\begin{enumerate}
\item A vortex might preserve the same (relative) velocity (with
the fluxoids) in both the above two regions, throughout a pinning
cycle. Hence, the travel times, for the two regions, would be
linearly proportional to their sizes. Namely,
\begin{eqnarray}
   {t_1 \over t_2}  & = & {l_1 \over l_2} = { d_{P} \over d_{\rm f} -
   d_{P}}, \ \ {\rm and  }\\
  {t_1 \over T_{\rm cyc}} & = & { d_{P} \over d_{\rm f}}
\end{eqnarray}
which results in
\begin{eqnarray}
F_{\rm pin} & = & {E_{\rm P} \over d_{\rm f}^2}    \\
          & \sim & 6 \times 10^{12} \, ({\rm dyn \ cm}^{-1}) \ E_{\rm MeV} B_{12}
\end{eqnarray}
where  $E_{\rm MeV}$ is the value of $E_{\rm P}$ in units of MeV.

\item Or, the vortex might be expected to adjust its velocity
instantaneously as it moves through the pinning and the free
regions, successively. The vortices might therefore spend almost
all of the time inside the pinning regions and fly across the free
spacings between the fluxoids rapidly, in almost zero time. That
is
\begin{eqnarray}
  t_2  & = & 0, \ \  {\rm and}\\
  t_1  & = & T_{\rm cyc}
\end{eqnarray}
which results in
\begin{eqnarray}
F_{\rm pin} & =   &   {E_{\rm P} \over d_{\rm f} d_{\rm P}} \\
                     & \sim & 1.4 \times 10^{16} \, ({\rm dyn \ cm}^{-1}) \ E_{\rm
                     MeV} \ d^{-1}_{12}
\end{eqnarray}
where $d_{12}$ is the value of $d_{\rm P}$ in units of
$10^{-12}$~cm.
\end{enumerate}
The above two cases might each represent a better approximation
for either a transient relaxation, following a sudden change of
the rotation frequency of the fluxoids (as for a glitch) or during
an equilibrium steady state spin-down phase, respectively. In
either case, the force is {\em independent} of the relative
velocity of the two families of lines; a dynamical feature
different than that of, say, the force due to the electron
scattering.

The corresponding vortex relaxation time $T_{\rm P}$, needed for a
given relative velocity between the fluxoids and vortices to be
dissipated, may be defined, from the EOM of a unit volume of the
charged component gas co-rotating with the fluxoids (the
``crust''), through
\begin{eqnarray}
n_{\rm v} {\vec {F}}_{\rm pin} & = &  \rho_{\rm c} {{\vec{v_{\rm
L}} - \vec{v_{\rm c}}}  \over T_{\rm P} }
\end{eqnarray}
where  $\vec{v_{\rm c}}$ is the local velocity of the electron gas
and the fluxoids, ie. the  ``crust'',  and $\rho_{\rm c}$ is its
effective density, corresponding to the contribution of the real
crust together with the free protons (and the electrons) in the
moment of inertia of the star. Note that since $F_{\rm pin}$ is a
velocity-independent force, $T_{\rm P}$ is the total time for the
decay of the velocity, and not an exponential time constant. For
an assumed jump $\Delta \Omega_{\rm c}$ in the rotation frequency
$\Omega_{\rm c}$ of the crust, out of its steady state co-rotation
with the vortices, one derives, from Eq.~16 together with Eq.~11
or 15,
\begin{eqnarray}
T_{\rm P} & = & 830 \, ({\rm s})\ {\Delta \Omega_{\rm c} \over
\Omega_{\rm c} } { \rho_{13} \over E_{\rm MeV} B_{12} }, \ {\rm or} \\
          & = & 0.4 \, ({\rm s})\ {\Delta \Omega_{\rm c} \over
\Omega_{\rm c} } { \rho_{13} d_{12} \over E_{\rm MeV} },
\end{eqnarray}
for the two approximations considered above, respectively, and
where $\rho_{13}$ is the value of $\rho_{\rm c} $ in units of
$10^{13} \ {\rm g \, cm}^{-3} $. Notice that the dependence on the
distance $r$ from the rotation axis  ($v = r \Omega$) has been
averaged out, and the radial velocity of the vortices, during a
relaxation, would have negligible effect.

Substituting typical values of ${\Delta \Omega_{\rm c} \over
\Omega_{\rm c} } \sim 10^{-6}$, and $ E_{\rm MeV} \sim \rho_{13}
\sim B_{12} \sim 1$, one finds, from Eq.~17, $T_{\rm P} \sim
10^{-4}$~s, which is much shorter than the corresponding time
scale due to the electron scattering $\tau_{\rm e} \sim
1$~--~$2$~s \citep{als84}. The estimated value of $T_{\rm P}$ for
the other case, from Eq.~18, is even smaller, by more than three
orders of magnitudes. Nevertheless, because $T_{\rm P}$ does
depend on the initial value of the induced relative velocity it
may, in principle, become larger than $\tau_{\rm e}$, but that
might happen only for the very large assumed disturbances, much
larger than that observed even in the giant glitches of the Vela
pulsar. For the steady state spin down, on the other hand, $T_{\rm
P}$ would be even much smaller than the above estimate.

The drag force, ${\vec {F}}_{\rm elec}$, of the electron
scattering off neutron vortices is, likewise, associated with a
vortex relaxation time $\tau_{\rm e}$, through an equation similar
to Eq.~16, as is the relation between the overall relaxation time
scale $\tau_v$ with the total force \( {\vec {F}}_{\rm tot} =
{\vec {F}}_{\rm pin} + {\vec {F}}_{\rm elec} \), namely
\begin{eqnarray}
n_{\rm v} {\vec {F}}_{\rm tot} & = &  \rho_{\rm c} {{\vec{v_{\rm
L}} - \vec{v_{\rm c}}}  \over \tau_v },
\end{eqnarray}
which means
\begin{eqnarray}
{1 \over \tau_v} & = & { 1 \over T_{\rm P} } + { 1 \over \tau_{\rm
e} }
\end{eqnarray}
Therefore, the frictional pinning force turns out to be the
dominant coupling mechanism for the core superfluid, and $ \tau_v
\sim T_{\rm P} $ may be adopted, in general.
\section{Coupling Timescale of the Superfluid}
The dynamical timescale for the rotational coupling of the bulk
superfluid to its environment (the crust) is the time needed for
the simultaneous re-adjustment of the vortices in both radial and
azimuthal directions in response to an external torque on the
superfluid, exerted primarily on the vortices. The external forces
on the vortices could, in general, be of a viscous drag or a
``static'' frictional nature \citep{adam85,jon91b}. For the
superfluid in the core of a neutron star, both types are present
and correspond to the electron scattering and the fluxoid
``scattering'', discussed above, respectively. The latter type,
associated with the ``pinning'' forces, should not be however
confused with the role of the pinning forces on the pinned
vortices co-rotating with the pinning centers. In order for the
pinning forces to act as frictional forces and impart a net torque
on the superfluid the vortices should maintain a {\em radial}
velocity, hence unpinning continuously due to the effect of the
Magnus force, which requires $|\omega| \geq \omega_{\rm cr}$
\citep{adam85,mj05a}. For $|\omega| < \omega_{\rm cr}$, random
unpinning events might play a role in the long-term coupling of
the superfluid, but the effect is neglected here, for simplicity
and more so because of its negligible effects on the quick
post-glitch responses. A more general treatment should include the
superfluid coupling rate driven also by the random unpinning of
the vortices \citep{mj06}, as has been already studied, in the
context of the vortex creep model, and found to be unimportant
\citep{sa09}.

\subsection{The case of free vortices (no pinning barriers)}
The dynamical timescale $\tau_{\rm free}$ of the superfluid
rotational relaxation, in the absence of any pinning of the
vortices  (Fig.~1a), has been previously determined
\citep{as88,mj98} as
\begin{eqnarray}
\tau_{\rm free} =  \ {I_{\rm s} \over I} \ \tau_{\rm e}  + {I_{\rm
c}^2 P_{\rm s}^2 \over 16 \pi^2 I_{\rm s} I} \  {\tau_{\rm
e}}^{-1}
\end{eqnarray}
This is the exponential timescale that appears in the solutions
obtained for the relaxations of the radial $r_{\rm v}(t)$ and the
azimuthal $\phi_{\rm v}(t)$ components of the vortex positions, in
polar coordinates on the equatorial plane, as a function of time
$t$. The time behavior of the vortex position (following an
assumed sudden rise in the rotation frequency of the ``crust'',
say at a glitch) is governed by the vortex equation of motion
(Eq.~1), using $\vec{F}_{\rm ex}= {\vec{F}}_{\rm elec}$, in the
absence of any pinning. The solutions, which were not given in the
correct form previously, are:
\begin{eqnarray}
r_{\rm v}(t) & = & r_0 \left[ {{\Omega_{\rm s}}_0 \over
{\Omega_{\rm F}} } + \left( 1- {{\Omega_{\rm s}}_0 \over
{\Omega_{\rm F}} } \right) e^{-t / \tau_{\rm free}} \right]^{1/2} \\
\phi_{\rm v}(t) & = & \phi_0 + {\Omega_{\rm F}} t + \left( {
I_{\rm s} \kappa n_{\rm v} \over I} \ \tau_e  - { I_{\rm c} \over
I \kappa n_{\rm v}  } \ {\tau_e}^{-1} \right) \ln{\left( {r_{\rm
v}(t) \over r_0 } \right) }
\end{eqnarray}
where 0-subscripts indicate initial values at $t=0$ of the
corresponding quantities, $I = I_{\rm c} + I_{\rm s} $, and
${\Omega_{\rm F}} = {1 \over I} ( I_{\rm c} {\Omega_{\rm c}}_0 -
I_{\rm s} {\Omega_{\rm s}}_0) $ is the final equilibrium frequency
of both the crust and the superfluid.
\subsection{The case of Pinned Superfluid}
For a pinned superfluid, already in a steady-state with \(
\Omega_{\rm c} = \Omega_{\rm L} < \Omega_{\rm s} \) and \( \omega
= \omega_\infty = \omega_{\rm cr} > 0 \), as discussed in \S~2.1,
a sudden jump in $\Omega_{\rm c}$ is then assumed to occur, the
cause of which is irrelevant for the purpose of the present
discussion. The crossing through the fluxoids by the vortices is
inevitable until the state of co-rotation between them is reached
(considering rigid vortex lines), hence the vortices would be
subject to forward azimuthal pinning forces. The response of the
pinned superfluid to such disturbances is argued, below, to be
{\em qualitatively} different, depending on the magnitude of the
increase $\Delta \Omega_{\rm c}$ in the rotation frequency of the
crust. The effect, should not be however attributed to the
particular form(s) of the pinning force, calculated above. Indeed,
the same types of behaviors would be expected even in the presence
of the drag force due to the electron scattering alone, as long as
the neutron vortices are assumed to be subject to the pinning to
the ``crust'' (co-rotating with the fluxoids and the charged
components). Also, note that the decomposition of the effect into
a ``jump'' in $\Omega_{\rm c}$ and a ``subsequent response'' of
the superfluid is not meant to be in real time, rather it is for
the sake of the analysis, as is common.
\subsubsection{Large Jump}
Following an assumed ``large'' jump, such that ${ \Delta
\Omega_{\rm c}}_{0} > \omega_{\rm cr}$ (Fig.~1c), the fluxoids
(and the electron gas) would be rotating faster than {\em both}
the vortices and the superfluid itself. Hence, the existing
forward azimuthal forces on the vortices would be consistent with
a corresponding {\em inward} motion of the vortices required for a
spin-up torque to be imparted to the superfluid. The steady state
lag would be thus washed out during (the jump and) the following
quick relaxation of the superfluid (and the vortices), bringing
the superfluid, first, to a state of co-rotation with the fluxoids
and the vortices. Hence, following a large jump the superfluid
will be, first, spun up by the crust, on a dynamical time scale
$\tau_{\rm pin}$, until the equilibrium state $ \Omega_{\rm s} =
\Omega_{\rm L} = \Omega_{\rm c}$ is reached (see Fig.~1c).

The coupling timescale $\tau_{\rm pin}$ may be determined from a
calculation similar to the above case of the free vortices, except
that $\tau_v$ (Eq.~20) would now be the relevant vortex relaxation
time. That is (compare with Eq.~21)
\begin{eqnarray}
\tau_{\rm pin}  =  \ {I_{\rm s} \over I} \ \tau_v  + {I_{\rm c}^2
P_{\rm s}^2 \over 16 \pi^2 I_{\rm s} I}  \  {\tau_v}^{-1}.
\end{eqnarray}
It may be noted that, the pinning force, $f_{\rm P}$, at each
intersecting point, is directed solely along the direction
perpendicular to the intersecting fluxoid; the component of the
force along the fluxoid being zero \citep{sau89,sa09}. In solving
the vortex equation of motion (Eq.~1), with the substitution
$\vec{F}_{\rm ex}= {\vec{F}}_{\rm pin}$, a decomposition of the
force to the radial and azimuthal components would be initially
faced with a difficulty, that may be however bypassed. The
projections of the fluxoids, on the equatorial plane, form an
array of parallel lines. In the presence of a difference between
the rotation rates of the vortices and the fluxoids, the vortices
at any given radial distance $r$ cut through those lines at
varying angles. Hence, the azimuthal component of the pinning
force, at each intersection, would vary between its maximum vale
$f_{\rm P}$ down to zero, for the vortices located at the
different azimuth angles. The same would be true for the radial
component of the force, as well. Thus, at any given time and for
the vortices at any given radial distance, the {\em average}
values of the azimuthal and the radial components of the pinning
force, per intersection, may be used; averaged over the vortices
at varying azimuthal angles over a circle. The average values,
would be the same for both components and, simply amount to $ {2
\over \pi} \, f_{\rm P}$. This will correspond to a change in the
relaxation time $T_{\rm P}$ by a factor ${ \pi \over 2}$, that we
have neglected.
\subsubsection{Small Jump}
In contrast, following an assumed ``small'' jump, in the rotation
rate of the crust, such that ${ \Delta \Omega_{\rm c}}_{0} <
\omega_{\rm cr}$ (Fig.~1b), there would follow {\em no transfer of
angular momentum} between the crust and the  superfluid, initiated
by the jump. The superfluid could be neither spun down nor up, by
the crust! Following the jump, the vortices would be indeed spun
up to come into a co-rotation with the fluxoids, due to the
existing forces. Nevertheless, the superfluid which is already
rotating faster than the crust, and its own vortices, (Fig.~1b)
could not be possibly spun up, and gain angular momentum from the
slower component, during such a spin-up of the vortices. Also, the
superfluid could not be {\em spun down}, during such a spinning up
of the crust, that is accompanied by a {\em spinning up} of the
vortices. Because, a spin-down of the superfluid requires a
radially {\em outward} motion of the vortices. Nonetheless, an
outward motion of the vortices may be realized only if they do
rotate, or tend to be rotating, faster than the the superfluid
environment, so that a slowing down {\em torque} could be imparted
on them (see \citet{mj05a,mj05b} for a more extended explanation;
indeed false conclusions, about the behavior of the superfluid in
the crust of a neutron star, have been communicated because of the
neglect of this primitive but fundamental fact, as discussed
therein).

Hence, following a small jump, the co-rotation of the vortices
with the crust is achieved with {\em no transfer} of angular
momentum to the superfluid, ie. at constant $ \Omega_{\rm s}$. The
time scale involved would be only that of the azimuthal vortex
velocity relaxation, $T_{\rm P}$. The speeding up of the vortices
alone does not, however, require any transfer of angular momentum,
since these are but massless (super)fluid configurations
(neglecting the inertia of the vortices, as is usually assumed).

\section{Discussion and Observational Implications}
Glitches are observed in radio pulsars as sudden changes $\Delta
\Omega_{\rm c} $ in the rotation frequency $\Omega_{\rm c}$ of the
crust with observed values of the jump in the range \( 10^{-9}
\lesssim \frac{\Delta \Omega_{\rm c}}{\Omega_{\rm c}} \lesssim
10^{-6} \).  In younger pulsars, the jump in $\Omega_{\rm c}$ is
also accompanied by an increase $\Delta \dot \Omega_{\rm c}$ in
the observed spin-down rate $\dot \Omega_{\rm c}$ of the crust,
with typical values of ${\Delta \dot \Omega_{\rm c} \over \dot
\Omega_{\rm c}} \lesssim 2.5\% $ \citep{Lyn87,kraw03}. The effect,
as such, has been explained successfully in terms of a decoupling
of the superfluid in the {\em crust} of a neutron star, which has
a fractional moment of inertia of similar magnitudes
\citep{alet84}. However, there has been observed cases with
\begin{eqnarray}
{\Delta \dot \Omega_{\rm c} \over \dot \Omega_{\rm c}} > 10\%
\end{eqnarray}
and recovery timescales up to $\sim 44$~d  \citep{Lyn87,Fln95}.
These observed cases would necessarily imply that part of the star
with the same fractional moment of inertia( ie. $> 10\%$, and
indeed up to $60\%$ at some observations) has been decoupled from
the crust, at the time of the observation. This, by itself, is a
definite proof of a rotational decoupling of (a part of) the {\em
core} of the star, at those glitches, since the crust does not
constitute that much of the moment of inertia of the star. Given
also the observed recovery time scales at these events, the
effect, in turn, indicates that a {\em pinned superfluid}
component of the stellar core must be involved, as has been argued
previously \citep{mj05b}. The core superfluid, in the {\em
absence} of pinning, does similarly cause observed large values of
$\Delta \dot \Omega_{\rm c} \over \dot \Omega_{\rm c}$, but that
would last only over a time of the order of $\tau_{\rm free}$
(Eq.~21). In contrast, the pinned superfluid remains decoupled
until the superfluid rotational lag, which is washed out quickly
at the jump (see Figs~ 1b \& 1c), recovers its steady state value
$\omega_{\rm cr}$. During this recovery period $t_{\rm PG}$, the
observable crust is being spun down, by the unchanged external
torque $N_{\rm ext}$, at an increased post-glitch rate $\dot
\Omega_{\rm PG}$, that could be as large as,
\begin{eqnarray}
\dot \Omega_{\rm PG} = {I \over I_{\rm c}} \, \dot \Omega_{\rm
SS},
\end{eqnarray}
where $\dot \Omega_{\rm SS} = {N_{\rm ext} \over I}$ is the steady
state value of the spin down rate of the star. Since, the vortices
are pinned and co-rotating with the fluxoids and the crust, thus
the superfluid is decoupled and the lag is built up due only to
the resulting decrease in $\Omega_{\rm c}$. Hence, the post-glitch
recovery timescale $t_{\rm PG}$, over which the increased
spin-down rate of the crust may persist, is expected to be
\begin{eqnarray}
t_{\rm PG}  =  \frac{\omega_{\rm cr}}{\dot \Omega_{\rm PG}} \,
\gtrsim 10^5 \ ({\rm s}),
\end{eqnarray}
where the estimated value is for an assumed low value of
$\omega_{\rm cr} = 10^{-4} \, {\rm rad \, s}^{-1}$, a large value
of ${I \over I_{\rm c}} = 20$, and a typical value of $\dot
\Omega_{\rm c} = 10^{-10} \, {\rm rad \, s}^{-2}$. Obviously,
larger recovery times are expected for the lower values of $\dot
\Omega_{\rm PG}$, that would be expected if only a fraction of the
core superfluid takes part in the decoupling-coupling mechanism.
Therefore, the largest observed values for the post-glitch
spin-down rates of the pulsars, as well as the associated recovery
timescales recorded, find a natural explanation in terms of the
expected decoupling of the superfluid component in the core, being
subject to the pinning with the fluxoids.

Further, the two cases of the large versus the small jumps,
considered above, would have different implications for the
observable post-glitch recovery. In the case of a large jump, the
expected initial spin-up of the core superfluid, discussed above,
would show up as a further increase, by an additional factor ${I
\over I_{\rm c}}$, in the observable spin-down rate of the crust,
compared to that given in Eq.~26. In contrast, no such an increase
is expected for the small jumps (compare Fig.~1b with 1c). Such a
difference in behavior, between the two cases, would however
persist only over time periods $ \sim \tau_{\rm pin}$, which makes
it indeed hard to be detected, observationally.

On the other hand, the two cases might be judged and distinguished
based also on the {\em earliest} detected values of $ \Delta
\Omega_{\rm c}$, as {\em compared} to the value of $\omega_{\rm
cr}$. That is, the two cases of small versus large jumps
correspond to (initial) values of $ \Delta \Omega_{\rm c} <
\omega_{\rm cr}$ and $ \Delta \Omega_{\rm c} > \omega_{\rm cr}$,
respectively (as indicated in Figs~1b and 1c). Notice that, for a
large jump  the condition $ \Delta \Omega_{\rm c} > \omega_{\rm
cr}$ would be still true, even after the initial relaxation of the
superfluid, over the time $\tau_{\rm pin}$, ie. at the post-glitch
$t = 0$ on Fig.~1c. Also, the observational uncertainty in the
exact epoch of the glitch, being of the order of few minutes,
would not blur between the two cases, since the change in $
\Omega_{\rm c} $ over such periods would be negligible. Therefore,
given the generally observed initial values for $\frac{\Delta
\Omega_{\rm c}}{\Omega_{\rm c}}$, cited above, as compared to the
expected values of $\omega_{\rm cr} \gtrsim 10^{-4} \ {\rm rad \
s}^{-1}$ for the pinned core superfluid, one may conclude that the
case of a small jump is the relevant one for most, if not all, of
the observed glitches.

Moreover, the two cases would have different implications also for
the possible cause of the glitches. A large jump has to be
necessarily induced by some ``external'' agent, other than the
core superfluid itself, simply because a donor of the angular
momentum could not spin up its counterpart to a frequency more
than its own. In contrast, and by the same token, a glitch induced
by the core superfluid would necessarily rank as a small one. The
core superfluid is thus, statistically, more favored as the
potential cause of the small jumps. The above mentioned indication
of the data for the prevalence of the small jumps, would thus
further support the glitch inducing mechanisms driven by the core
superfluid \citep{mt85,mj02}.

Finally, the pinned core superfluid  does not respond to a jump in
$\Omega_{\rm c}$, at a glitch, smaller than the critical value of
the superfluid rotational lag, $\omega_{\rm cr}$, as indicated
above. The angular momentum gained by the crust, at such glitches,
would not be shared, any further, with the core superfluid. The
case is highlighted here again, not because of any prominent
observable effect different than the other case of the large
jumps. Rather, for its prominent conceptual value in demonstrating
the vital requirement for the presence of a corresponding torque
for a spin-down (-up) of a superfluid to be achieved. The motion
of the superfluid vortices too has to conform with this basic
physics, be it a smooth motion in the absence of any pinning, or
the so-called creeping of the pinned vortices.

This work was supported by a grant from the Research Committee of
Shiraz University.

\appendix

\section{The average rate of crossing with fluxoids, per unit length of a vortex}
We are assuming the fluxoids, in the core of a neutron star, form
a uniform array of parallel lines, along the magnetic axis
($\vec{B}$) of the star, having a total number of $N_{\rm f}$,
within a spherical boundary surface of radius $ R$. Likewise, the
neutron vortices constitute a uniform lattice of parallel lines,
along the rotation axis ($\vec{\Omega}$) of the star, having a
total number of $N_{\rm v}$, within the same spherical boundary.
The two axes are in general inclined at an angle $\chi$, of which
we consider first the case with $\chi = {\pi \over 2}$, for
further clarity, and will generalize at the end. Taking the z-axis
along $\vec{\Omega}$, the fluxoids are thus parallel to the x-y
plane.

\begin{enumerate}
\item A circular slab, perpendicular to the z-axis, its rim at the
polar angle $\theta$, a distance $z=R \cos{\theta}$ from the
origin, with a radius $R_{\rm z} = R \sin{\theta}$, and a
thickness $dz = \sin{\theta} \, d\theta$, includes a number
$N_{\rm b}$ of fluxoids, being parallel to it, where
\begin{eqnarray}
N_{\rm b} & = & {N_{\rm f} \over \pi R^2} \ (2 R_{\rm z} \, dz) \\
          & = & {2 \over \pi} \ N_{\rm f} \ {\sin^2{\theta}} \ d\theta
\end{eqnarray}
\item The fraction of the fluxoids in the slab, lying at
(cylindrical) distances $r$ to $ r + dr$ from the z-axis is
\begin{eqnarray}
N_{\rm s}(r) & = & {dr \over 2 R_{\rm z} } \, N_{\rm b}
\end{eqnarray}
\item Each of the lines at $r$, sweeps an annular area $s(r)$, in
the x-y plane, per rotation cycle, that is
\begin{eqnarray}
s(r) = 2 (\pi {R_{\rm z}}^2 - \pi r^2)
\end{eqnarray}
\item The area $dS_{\rm b}$ swept by all the fluxoids in the slab
would be
\begin{eqnarray}
dS_{\rm b}  & = & 2 \int_{0}^{R_{\rm z}} N_{\rm s}(r) \,  s(r) \\
           & = & {4 \over 3} \, \pi {R_{\rm z}}^2  \, N_{\rm b}
\end{eqnarray}

\item The cumulative total area $S_{\rm tot}$ swept by all the
fluxoids within the sphere, parallel to the x-y plane, thus
becomes
\begin{eqnarray}
S_{\rm tot} & = & 2 \int_{\theta = 0}^{\pi \over 2} dS_{\rm b} \\
          & = & \pi {R}^2  N_{\rm f}
\end{eqnarray}
This may be compared with a corresponding value $ {4 \over 3} \pi
{R}^2 N_{\rm f} $ which would result for an assumed cylindrical
geometry, or with a value $ 2 \pi {R}^2 N_{\rm f}$ if all the
fluxoids were of the same length $2 R$, lying radially.

\item The number $ N_{\rm X} $ of crossings between the fluxoids
and the vortices (the latter all passing through the x-y plane,
with a number density $n_{\rm v} = {N_{\rm v} \over  \pi {R}^2}$),
per cycle of relative rotation, per unit area of the x-y plane,
becomes
\begin{eqnarray}
 N_{\rm X} & = & n_{\rm v} \, S_{\rm tot} = N_{\rm v} \, N_{\rm f}
\end{eqnarray}
which may be, in turn, compared with a corresponding value $2 \,
N_{\rm v} \, N_{\rm f}$, if each fluxoid were to cross each and
any of the vortices, per relative rotation cycle. \\
\item The total sum  \( L_{\rm v}\) of the lengths of all the
vortices, within the sphere, is
\begin{eqnarray}
 L_{\rm v} & = & n_{\rm v} \, ({4 \over 3} \, \pi \, R^3) = {4 \over
 3} \, R \,  N_{\rm v}
\end{eqnarray}
Hence, the average number \( N_1 \) of crossings, per unit length
of a vortex, per rotation cycle, is
\begin{eqnarray}
N_1  = \frac{N_{\rm X}}{L_{\rm v}} = {3 \over 4} \, \frac{N_{\rm
f}}{R}
\end{eqnarray}
\item For an square lattice of the fluxoids, with a spacing
$d_{\rm f}$, between the nearest fluxoids,
\begin{eqnarray}
N_{\rm f} = \frac{\pi \, R^2}{d^2_{\rm f}}
\end{eqnarray}
Thus,
\begin{eqnarray}
N_1 = {3 \over 4} \, \frac{\pi \, R}{d^2_{\rm f}}
\end{eqnarray}
\item The point of intersection of each vortex with the x-y plane,
at a distance $r$ from the z-axis, moves a distance  $d(r) = 2 \,
\pi \, r$, per rotation cycle. The corresponding average distance
\( d_{\rm v}\) that a vortex moves, becomes
\begin{eqnarray}
d_{\rm v} = {1 \over R} \, \int_{0}^{R} d(r) \, dr = \pi \, R
\end{eqnarray}
\item Therefore, the rate \( N_{\rm c} \) of crossings, per unit
length of a vortex, per pinning cycle, is derived as
\begin{eqnarray}
N_{\rm c} = {d_{\rm f} \over d_{\rm v}} \, N_1 = {3 \over 4} \, {1
\over d_{\rm f}}.
\end{eqnarray}
\item For the general case of an arbitrary inclination angle
$\chi$, the derivation  would be similar, except that one starts
by considering a slab, parallel to the fluxoids and at the same
angle $\chi$ with respect to the rotation axis, so that it
preserves its angle during the rotation. Then the projected area
swept by those fluxoids, on the x-y plane would have an additional
factor $\ \sin{\chi}$, as compared to that in Eq.~A8. As a
consequence, the total number of the crossings would be corrected
by the same factor, namely $N_{\rm X} = N_{\rm v} \, N_{\rm f} \,
\sin{\chi}$, and also for the final result, $ N_{\rm c} = {3 \over
4} \, {1 \over d_{\rm f}} \, \sin{\chi} $.

\end{enumerate}

\begin{figure}
\caption{Schematic representation of the relative values of the
angular velocities of the superfluid $\Omega_{\rm s}$ ({\em
dotted} line), the vortices $\Omega_{\rm L}$ ({\em dashed} line),
and the ``crust'' $\Omega_{\rm c}$ ({\em full} line), excluding
the overall spinning down of the system. On each plot, three
successive phases are shown at a relaxation, namely an earlier
steady state (marked as ``S.S.''), an assumed sudden increase in
$\Omega_{\rm c}$ (``JUMP'') and the subsequent relaxed state
(``POST-G''). {\bf a)} is for the case of free vortices, in the
absence of pinning barriers, whence the steady state spinning down
corresponds to (an almost) co-rotation of the vortices, with the
superfluid and the crust. The assumed increase in the angular
velocity of the crust is shared with the core superfluid over its
dynamical timescale $\tau_{\rm D} = \tau_{\rm free}$. {\bf b)} and
{\bf c)} are both for a pinned superfluid. Due to the assumed
pinning with flux lines, in the initial spinning-down
steady-state, a rotational lag between the superfluid and the
vortices is necessary. For the ``small'' jumps, as in {\bf b)},
only the vortices are spun-up over an ``azimuthal'' relaxation
timescale ${T_{\rm P}}_\phi = T_{\rm P}$. For the ``large'' jumps,
as in {\bf c)}, the pinned core superfluid couples to the crust
over its corresponding dynamical timescale $\tau_{\rm D} =
\tau_{\rm pin}$, as discussed in the text.}
\end{figure}

\newpage

\begin{figure}[]
\vbox to26.cm{\rule{0pt}{24.cm}} \includegraphics{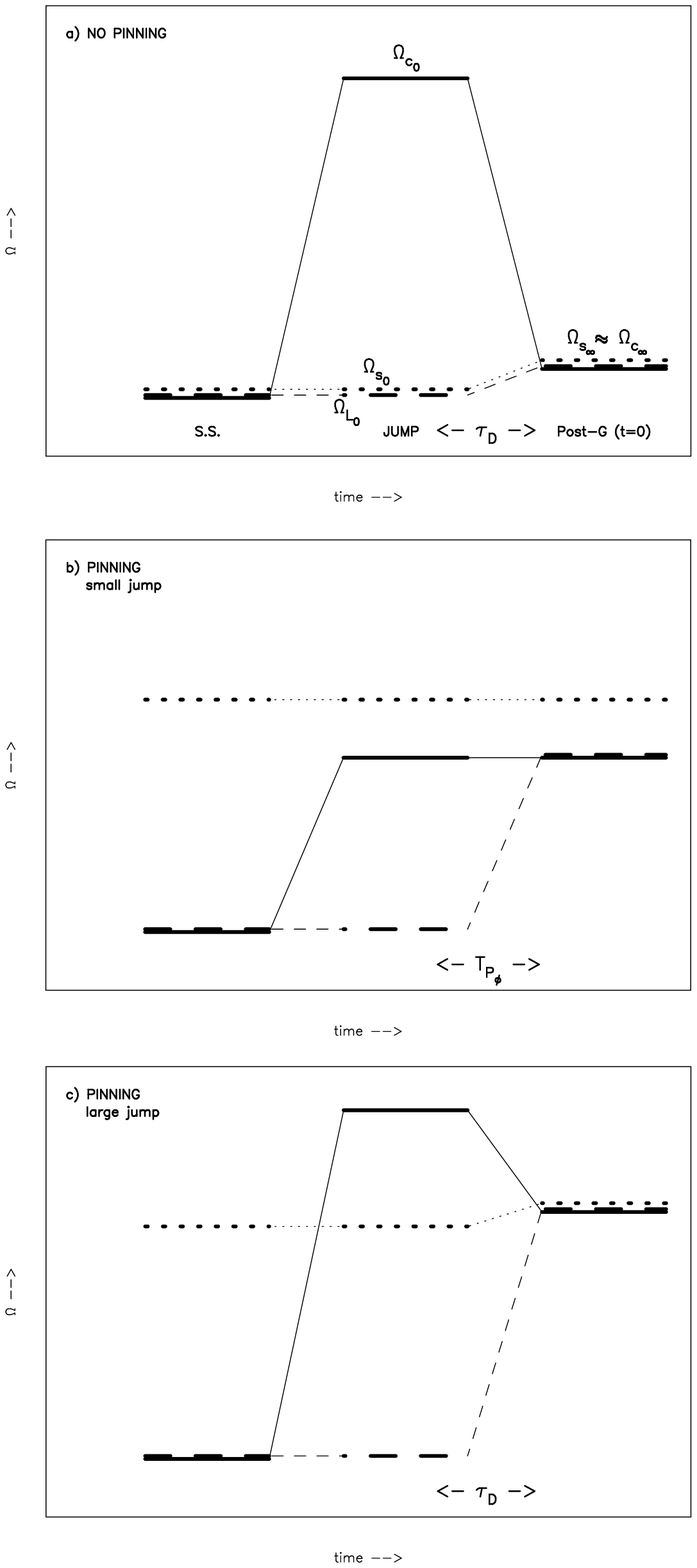}
\end{figure}
\end{document}